\documentclass[12pt]{article}
\usepackage{etex}
\usepackage{mathpazo} % math & rm

\usepackage[scaled]{helvet} % ss
\usepackage{courier} % tt
\normalfont
\usepackage[T1]{fontenc}
\usepackage{authblk}
\usepackage{float}
\usepackage{multirow}
\usepackage[margin=0.5in]{geometry}
\usepackage{subfigure}
\usepackage[subfigure]{tocloft}
\usepackage{appendix,graphicx,epstopdf,mdwlist,enumerate,mathrsfs,amsmath,amssymb,verbatim, fancyhdr, amsfonts}
\usepackage[nottoc,numbib]{tocbibind} %This will include the Bibliography in the Table of Contents with numbering
\usepackage[singlespacing]{setspace}
\usepackage[bottom]{footmisc}
\usepackage{indentfirst}
\usepackage{endnotes}
\usepackage{lscape}
\usepackage{multicol}
\usepackage{sectsty}
\usepackage{adjustbox}
\usepackage{titlesec} %[small]
\usepackage{footnote}
\usepackage{booktabs}
\usepackage{booktabs}
\usepackage{sgame,tikz}
\usetikzlibrary{external,automata,trees,positioning,shadows,arrows,shapes.geometric,calc, matrix,decorations}

\usepackage{ctable}
\usepackage{microtype}%[longnamesfirst]
\usepackage{natbib}
\usepackage{titling}
\usepackage{abstract}
\usepackage{blindtext}
\usepackage[flushleft]{threeparttable}
\usepackage{adjustbox}
\usepackage{resizegather}
\usepackage{placeins}
\usepackage{lscape}
\usepackage{caption}
\usepackage{geometry}
\usepackage{tocloft}
\usepackage{sectsty}
\usepackage{hyperref}
\usepackage{enumitem}

\makeatletter
\def\@biblabel#1{\hspace*{-\labelsep}}
\makeatother 
\renewenvironment{abstract}
{\centerline{\bf Abstract}
\begin{list}{}
{\setlength{\rightmargin}{0.9cm} \setlength{\leftmargin}{0.9cm}}
\item[]\ignorespaces}
{\unskip\end{list}}
\renewcommand{\baselinestretch}{1.5}
\newcommand{\qed}{\nobreak \ifvmode \relax \else
      \ifdim\lastskip<1.5em \hskip-\lastskip
      \hskip1.5em plus0em minus0.5em \fi \nobreak
      \vrule height0.75em width0.5em depth0.25em\fi}

\renewcommand{\captionsize}{\normalsize}
        \sectionfont{\Large}
         \subsectionfont{\large}
\hypersetup{
    colorlinks=true,%
    citecolor=blue,%
    filecolor=green,%
    linkcolor=[rgb]{0.21,0.46,0.53},%
    urlcolor=[rgb]{0.27,0.51,0.71}
}   % useful for program listings
\begin{document}
\renewcommand{\baselinestretch}{1.5}
\singlespacing %

\setlength{\droptitle}{-3em}
%\vspace{-5cm} 

\begin{titlepage}

\thispagestyle{empty}
\title{Black-box Optimizers vs Taste Shocks\thanks{I would like to thank Leonardo Martinez and Emircan Yurdagul for their constructive comments and suggestions, and especially Bulent Guler for his incredible support, especially in the execution of taste shocks. The author acknowledges financial support from the Research Foundation–Flanders and Ghent University Starting Fund.}}
\author{Yasin K\"{u}r\c{s}at \"{O}nder\thanks{Deparment of Economics, Ghent University; email: kursat.onder@ugent.be}}
\affil{Ghent University}

%\date{}
\maketitle \
%\vspace{-1cm} 
\pagenumbering{roman}

\begin{centering}

\medskip
%\textbf{---------- PRELIMINARY AND INCOMPLETE-----------}\\
%\textbf{---------- DO NOT CIRCULATE-----------}

\end{centering}
\begin{abstract}
%\vspace{-1cm}
\begin{singlespace}
\noindent 
We evaluate and extend the solution methods for models with binary and multiple continuous choice variables in dynamic programming, particularly in cases where a discrete state space solution method is not viable. Therefore, we approximate the solution using taste shocks or black-box optimizers that applied mathematicians use to benchmark their algorithms. We apply these methods to a default framework in which agents have to solve a portfolio problem with long-term debt. We show that the choice of solution method matters, as taste shocks fail to attain convergence in multidimensional problems. We compare the relative advantages of using four optimization algorithms: the Nelder-Mead downhill simplex algorithm, Powell's direction-set algorithm with LINMIN, the conjugate gradient method BOBYQA, and the quasi-Newton Davidon-Fletcher-Powell (DFPMIN) algorithm. All of these methods, except for the last one, are preferred when derivatives cannot be easily computed. Ultimately, we find that Powell's routine evaluated with B-splines, while slow, is the most viable option. BOBYQA came in second place, while the other two methods performed poorly.
\end{singlespace}
\end{abstract}

\pagestyle{plain}
\newcounter {newcounter}
\end{titlepage}\setstretch{1.5}
\renewcommand{\baselinestretch}{1.5}
\pagenumbering{arabic}
\onehalfspacing

%\tableofcontents % insert the table of contents
%\newpage

\section{Introduction}
An increasingly common problem in dynamic programming is solving models with binary and multiple choice variables. In this study, we examine a class of models where the discrete state space (DSS) method is not applicable. To address this issue, we compare and extend solution methods using taste shocks and black-box optimization algorithms, which are commonly used by applied mathematicians to benchmark their algorithms. We intentionally apply these methods to a framework that includes long-term debt and default behavior for two main reasons. First, previous studies have shown that the DSS method is not a viable solution option for this framework (\cite{JIE} and CE \cite{Chatty}). Second, borrowing and default behavior are key characteristics of consumers, firms, and sovereigns.

Earlier literature has shown that both DSS and interpolation methods tend to deliver identical results for finer grids under one-period and one-asset models. However, this is not the case with long-term debt, as achieving convergence using only DSS, without incorporating additional factors such as taste shocks, appears to be extremely challenging, as previously discussed in CE. We evaluate and improve upon taste shock solution methods and black-box optimizers with alternative interpolation techniques to shed light on their practicality, as well as the bottlenecks that these solution methods entail. We then extend the baseline model to a portfolio allocation problem and demonstrate that the solution method used matters. Attaining convergence with taste shocks now becomes a near-impossible task, whereas interpolation methods with \textit{black-box} multidimensional search routines (as described in \cite{Numerical_Recipes}) can achieve that objective. Therefore, the comparison of different optimization algorithms is an important step in determining the best approach for solving a particular problem. Nelder-Mead downhill simplex algorithm, Powell's direction-set algorithm with LINMIN, and conjugate gradient method BOBYQA are all well-established methods that can be used when derivatives cannot be easily computed. These methods have been extensively tested and have been found to be reliable and efficient in a wide range of optimization problems.

However, the relative performance of these algorithms can vary depending on the specific problem at hand. Therefore, it is important to evaluate the performance of each algorithm on the particular problem being considered. This can be done by comparing the convergence speed, accuracy, and robustness of each algorithm. In the context of portfolio allocation problems, the choice of optimization algorithm can have a significant impact on the quality of the solution obtained. Therefore, it is important to carefully evaluate the performance of different algorithms on this problem.

When it comes to formulating long-term debt, both CE and HM use a similar approach. The idea is to avoid the need to track the distribution of bond maturities. However, each model has a slightly different method. CE offers a more comprehensive framework for long-term debt, of which HM is a particular case. The primary divergence arises due to the various definitions of duration proposed by researchers. Depending on the calibration, these definitions may lead to substantially different durations and spreads.

To solve the model, CE introduces a temporary income shock drawn from a continuous distribution to boost the convergence properties of their DSS method, while HM uses interpolation techniques. The new shock in the CE framework works as a taste shock to smooth out kinks. Importantly, this additional shock, which is introduced as a state variable, limits the researcher's experimentation to establish its role. We introduce it in a more innovative way to speed up the convergence and to avoid the limitation of memory problems. In particular, with taste shocks on the debt choice, conditional on the current debt choice and income, the debt policy becomes a density function as a function of the future debt choice, as opposed to being a singleton in the absence of taste shocks. Yet, this density function turns out to be positive for a handful of bond choices. Thus, we restrict our evaluations to the choice probabilities for the adjacent grid points around the maximum that attains the global maximum for the objective function, using sparse matrices.

The main issue with taste shocks arises when the grid size is increased: the model moments tend to change, albeit slightly. Therefore, when increasing the grid size, one needs to adjust by reducing the standard deviation of the taste shock to achieve convergence, which can be a tedious task.\footnote{We would like to thank Emircan Yurdagul for confirming this finding.} However, with interpolation techniques, researchers can use one-dimensional minimization routines, which can simplify the process. Among spline and linear interpolations, we find that linear interpolation is at least 20 times faster than its spline counterpart, while delivering nearly the same results. Nevertheless, linear interpolation may sometimes fail to achieve convergence. Therefore, we switch to spline interpolation after a certain number of iterations.

 The superiority of black-box optimizer over taste shocks becomes particularly notable when solving a portfolio problem. We extend the baseline model by introducing one-period non-defaultable debt and use black-box optimizers to solve the model. Among the three benchmark optimizers that do not require derivative computation, the Nelder-Mead downhill simplex algorithm and the conjugate gradient method BOBYQA are lagging behind the Powell's direction-set algorithm evaluated with splines. However, when introducing taste shocks, we failed to achieve convergence in a relatively simple portfolio problem due to the curse of dimensionality. This makes taste shock techniques significantly inferior to optimizer techniques when addressing portfolio problems. Despite this failure, we developed an algorithm that would speed up computation and avoid memory issues for one-asset problems.

\textbf{Literature Review:} \cite{mcfadden74} introduce taste shocks with a functional form known as the Generalized Extreme Value distribution in the context of discrete choice models and it was then extended into dynamics models in fields of industrial organization, structural labor and international trade (\cite{Pakes1986} and \cite{rust1987}) and extended into endogenous grid method (\cite{Iskhakov2017}). The same idea is also applied to the consumer and sovereign default settings under one-period debt by \cite{Chatty2023} and \cite{Emircan2021}. We extend on these studies on three main fronts: $(i)$ solving it with long-term debt, $(ii)$ providing an algorithm to limit the curse of dimensionality problem as well as to speed up the convergence by utilizing sparse matrices, and $(iii)$ documenting pitfalls of introducing taste shocks particularly when additional choice variable is introduced such as the portfolio problem introduced here.

\cite{Nota} compared the quantitative performance of sovereign default models with one-period debt by benchmarking against \cite{Arellano05} and \cite{AG_06}. They found that both the DSS and interpolation techniques converge to similar results when the number of DSS grid points is sufficiently high. However, we demonstrate that this is not the case with long-term debt. Although small, the approximation techniques produce different results, and the DSS technique alone is insufficient to achieve convergence. We extend our analysis by introducing a portfolio problem, which several other researchers have also explored by adding additional asset classes to standard long-term debt default models (\cite{Reserves}, \cite{RED_Greece}, \cite{ArellanoRamanarayanan}, \cite{Eurobonds}, \cite{Onderswap}, \cite{hidden}, \cite{gdp_indexed}). We contribute to this literature by comparing the performance of alternative black-box optimizers and solution methods.

The remainder of the paper proceeds as follows. Section \ref{sec:model} presents the model, and Section \ref{sec:calibration} explains the benchmark calibration. Section \ref{sec:results} features the simulation results. Finally, Section \ref{sec:conclusion} concludes.

\section{Model \label{sec:model}}
This section features a dynamic small open economy model with long-term non-contingent debt. The economy is endowed with a single tradable consumption good. The shocks to the economy are uninsured, and time is discrete.

\subsection{Environment}
\textbf{Preferences and endowment.} This paper considers a small open economy in which the benevolent government maximizes the utility of the infinitely lived representative household and has preferences given by
\begin {equation}
E_0\sum_{t=0}^{\infty}\beta^tu(c_t),
\end {equation}
where $E$ denotes the expectation operator, $0<\beta<1$ is the intertemporal discount factor, and $c_t$ denotes aggregate consumption at time $t$. The utility function $u(.)$ is an increasing, continuous, and strictly concave function that belongs to the class of CRRA utility functions and can be written as
\begin{equation}
u(c) = \frac{c^{1-\gamma}}{1-\gamma}
\end{equation}
with a risk aversion parameter $\gamma$.  Each period, households receive a stochastic income $y \in \mathcal{Y} \subset \mathbb{R}_{++}$ that follows an AR(1) process:
\begin{equation}
\log (y_{t})=\left( 1-\rho \right) \mu + \rho \log
(y_{t-1})+\varepsilon _{t}\label{eqn:AR1},
\end{equation}
with $|\rho |<1$, and $\varepsilon _{t}\sim N\left( 0,\sigma
_{\epsilon }^{2}\right)$.

\textbf{Asset space.} CE and HM use lightly different formulations of long-term debt. In the case of HM the government issues a perpetuity that pays a geometrically decreasing coupon $\kappa$ at a rate $\delta \in (0,1]$. If default is avoided, the price of the debt will be risk-free, and the average one-period bond price is equivalent to the present value of non-contingent long-term debt. Coupon payment $\kappa$ is calculated such that the default-free price of the debt is equal to the average one-period debt price. A non-contingent bond issued in period $t$ promises to pay ($\kappa, (1-\delta)\kappa, (1-\delta)^2\kappa,...,0$) units of the tradable good in subsequent periods. Hence, long-term debt dynamics can be represented as follows:
\begin{equation*}
b_{t+1} = (1-\delta)b_{t} + l_{t},
\end{equation*}
where $b_{t}$ and $b_{t+1}$ are the total number of outstanding non-contingent debt coupon claims at the beginning of time $t$ and $t+1$, and $l_{t}$ is the number of non-contingent long-term bonds issued in period $t$. Notice that $\delta = 1$ is a special framework of long-term bonds and corresponds to the economy with one-period debt and one can write $\kappa = \frac{\delta+r}{1+r}$ where $r$ is the risk-free interest rate and the risk-free price in HM setting equals to $\frac{1}{r}$.

In the CE world, the long-term debt is not perpetuity anymore, but debt now matures probabilistically. In particular, a unit of outstanding debt matures with a probability of $\delta$. The unmatured amount pays a coupon payment $z$ with a probability $(1-\delta)$. The risk-free price in the CE setting then becomes $\frac{\delta+(1-\delta)z}{r+\delta}$. Both studies intuitively use the same trick but with a slightly different approach. The intuition is to avoid tracking the entire distribution of different bond maturities between a government's default and repayment decisions.

\textbf{Timing.} The government starts a period with non-contingent debt $b$, and income $y$, which are all public information. The government then chooses to default, and if repaid, then lenders offer a menu of prices depending on the government's amount of new debt issuance and the current income.\footnote{This situation is identical to the government offering lenders a menu of prices and lenders choosing how much to lend in equilibrium. The majority of the studies in the quantitative default literature describe similar game setups.} The government then chooses its next period debt for its budget balances with which consumption reads using the HM formulation of long-term debt
\begin{equation}
c = y - \kappa b + q^{HM}(b^{\prime},y)(b^{\prime} - (1-\delta)b),\label{eq:c_HM}
\end{equation}
and using the CE formulation
\begin{equation}
c = y + (\delta+(1-\delta) z)b + q^{CE}(b^{\prime},y)(b^{\prime} - (1-\delta)b).\label{eq:c_CE}
\end{equation}

Notice that with $z=\frac{r}{1+r}$, HM formulation becomes a special version of the CE formulation, except the duration definition. In CE, the duration equals to $\frac{1}{\delta}$, whereas the definition of duration in \cite{Macaulay1938} is used in the HM world. This minor difference may yield different outcomes when a researcher calibrates his/her model.\footnote{Duration $D$ is the weighted average maturity of future cash flows. A bond issued at time $t$ that promises to make periodic payments $\kappa$ for the subsequent periods at time $1, 2, \ldots, J$ years into the future with the final price of zero has duration $D$. $D =\frac{1+i}{i+\delta}$ where $i$ is the periodic yield a lender would earn if the bond is held to maturity with no default and it satisfies
\begin{equation*}
q = \sum_{j=1}^{\infty}\frac{\kappa(1-\delta)^{j-1}}{(1+i)^{j}}.
\end{equation*}
The sovereign spread $r_s$ in HM economy is computed as the difference between yield $i$ and the risk-free rate $r$. The annualized spread reported in the tables is computed as \[1+r_s = \left(\frac{1+i}{1+r}\right)^4.\] The debt levels obtained from the simulations are equivalent to the present value of future debt obligations and computed as $\frac{b^{\prime}}{\delta+r}$.\label{footnote_spread}} Another minor reporting difference is HM divides $\frac{b}{y}$ by 4 to present the average debt-to-income ratio while CE does not.

If the government defaults on its debt, then a penalty scheme is imposed; the government is not allowed to borrow from international markets for a stochastic number of periods with an income cost of defaulting during its exclusion. Its consumption during exclusion is given by \[c = y - \phi(y),\] where default cost $\phi(y)$ is equal to $Max\{0,d_0y+d_1 y^2\}$. The government starts with zero debt when access to the credit markets is regained.

\subsection{Recursive formulation}
Let $V_{R}$, $V_{D}$ denote the value of repayment and the value of default functions where subscripts $R$ and $D$ represent the repayment and default, respectively. Let $V$ be the value function for the government that has the option to default. For any non-contingent price function $q$, $V$ satisfies the following functional equation:
\begin{equation}
V(b,y)=\underset{}{Max }\left\{ V_{R}(b,y),V_{D}(y)\right\},  \label{eq:V_CE}
\end{equation}
where the government's value of repayment is given by
\begin{eqnarray}
& & V_{R}(b,y)  =  \max_{c\geq0,\, b^{\prime}\in \mathcal{B}}\left\{u\left( c \right) +  \beta
\mathbb{E}_{y^{\prime}|y}  V(b^{\prime},y^{\prime }) \right\}, \label{eq:VR_CE} \\%
&&\text{subject to } \notag \\
c &=& y - (\delta+(1-\delta)z) b + q^{CE}(b^{\prime},y)(b^{\prime} - (1-\delta)b).\notag
\end{eqnarray}

The defaulting government regains access to the credit markets with a constant probability $\psi \in[0,1]$ and has zero debt at the time of market reentry. The value of defaulting reads
\begin{eqnarray}
&& V_{D}(y)  =  u\left( c\right) + \beta \mathbb{E}_{y^{\prime}|y} \left[(1-\psi)V_{D}(y^{\prime}) + \psi V(0,y^{\prime })\right], \ \ \label{eq:VD_CE} \\
&& \text{subject to } \notag \\
&& c = y - \phi(y). \notag 
\end{eqnarray}
The solution to this problem yields a binary default decision rule $\hat{d}(b,y)$ $\in$ $\{0,1\}$, 1 if the government defaults, and 0 otherwise, and a borrowing rule $\hat{b}(b,y)$. In equilibrium, defined in Section \ref{sec:equilibrium}, lenders use these decision rules to price contracts.
%\vspace{-0.5cm}
%\subsection{Investors' problem\label{sec:shareholder}}
Now it remains to describe the lenders' problem. The bond market is competitive, and lenders are assumed to be risk-neutral and atomistic. Thus, lenders take the price schedule $q(b^{\prime},y)$ as given. The opportunity cost of funds is given by the exogenous risk-free interest rate $r > 0$. With the zero-profit assumption and no-arbitrage condition in place, price functions for non-contingent bonds $q^{CE}$ solve the following functional equation:

%\fontsize{10}{15}{
\begin{eqnarray}
q^{CE}(b^{\prime},y)  &=& \frac{\mathbb{E}_{y^{\prime}|y}\left[\left( 1 - \hat{d}\left(b^{\prime},y'\right)\right) \left[\delta + (1-\delta)\left(z+ q^{CE}\left( b^{\prime\prime},y^{\prime}\right)\right)\right] \right]}{1+r}, \label{eq:q_CE}
\end{eqnarray}
%}
where $d^{\prime}$ = $\hat{d}\left(b^{\prime},y'\right)$ denotes the next-period equilibrium default decision, $b^{\prime\prime}$ $=$ $\hat{b}( b^{\prime},y^{\prime})$ denotes the next-period equilibrium non-contingent bond decision. 

In equation (\ref{eq:q_CE}), the expected value of holding a bond should be equal to the expected return of holding in a risk-free asset with a return of $r$. If a lender purchases non-contingent debt today, conditional on the government's repayment next period, the lender will receive $\delta + (1-\delta)z$ units of goods plus the unmatured amount of its receivables, which are worth $(1-\delta) q^{CE} \left( b^{\prime\prime}, y^{\prime}\right)$.

As noted before, with $z=\frac{r}{1+r}$, the HM formulation becomes a special version of the CE formulation, in particular, the budget constraint and the price function in the HM world reads as:
\[c = y - \kappa b + q^{HM}(b^{\prime},y)(b^{\prime} - (1-\delta)b),\]
%c = y - (\delta+(1-\delta)z) b + q^{CE}(b^{\prime},y)(b^{\prime} - (1-\delta)b)
\begin{eqnarray}
q^{HM}(b^{\prime},y)  &=& \frac{\mathbb{E}_{y^{\prime}|y}\left[\left( 1 - \hat{d}\left(b^{\prime},y'\right)\right) \left[\kappa + (1-\delta) q^{HM}\left( b^{\prime\prime},y^{\prime}\right)\right] \right]}{1+r}. \label{eq:q_HM}
\end{eqnarray}
Notice that in the absence of default risk, the price of the non-contingent long-term debt equals $q^* = \frac{\kappa}{r+\delta} = \frac{1}{1+r}$, so one can write $\kappa = \frac{r+\delta}{1+r}$. 
\subsection{Definition of equilibrium\label{sec:equilibrium}}
This paper focuses on a Markov perfect equilibrium (MPE). Hence, equilibrium conditions on the current state variables, not on the entire history.

\noindent \textbf{Definition 1 (Markov perfect equilibrium)}  A \textit{Markov perfect equilibrium} is characterized by
\begin{enumerate}%[I]
\item a collection of value functions $V$, $V_{R}$, and $V_{D}$;
\item rules for default $\hat{d}$ and borrowing $\hat{b}$; and
\item debt price function $q^{CE}$;
\end{enumerate}
such that:
\renewcommand{\theenumi}{\roman{enumi}}
\begin{enumerate}%[I]
\item given price function $q^{CE}$; $ \left\{V, V_{R}, V_{D}, \hat{d}, \hat{b} \right\}$ solve the Bellman equations  (\ref{eq:V_CE}),  (\ref{eq:VR_CE}), and (\ref{eq:VD_CE}).
\item given policy rules $ \left\{\hat{d}, \hat{b}\right\}$, the price function $q^{CE}$ satisfy equation (\ref{eq:q_CE}).
\end{enumerate}

\subsection{Taste Shocks}

\cite{Chatty} proposed adding taste shocks to the model to improve the solution of the model. They show that without taste shocks DSS method cannot achieve convergence in the pricing function with computationally feasible grid size for the state variables. CE introduces taste shocks as transitory endowment shocks with a small standard deviation. However, this limits the benefit of using the taste shocks since the taste shock becomes a state variable facing the curse of dimensionality and the expected continuation value requires the computation of an integral which complicates the solution further and adds another layer of computational inaccuracy. Next, \cite{Emircan2021} introduce the taste shocks following \cite{mcfadden74} and \cite{rust1987} without increasing the state space so that increasing grid size does not become a problem. Technically, one needs to show that convergence-satisfying taste shock does not change the model moments and one way of doing so is to solve the model by setting the std of the taste shock to zero.

The two main differences of the model with taste shocks are restricting the debt choice space to a finite discrete choices and the addition of taste shocks conditional on each discrete debt and default choice. Following \cite{mcfadden74} and \cite{rust1987}, we assume these taste shocks are additively separable and i.i.d Extreme Value type I distributed. We allow separate taste shocks for the discrete default choice and discretized debt choices. Given these changes, the value function for the government before the default choice becomes 
\begin{equation}
V(b,y,\epsilon)=\max\left\{ V_{R}(b,y)+\sigma_{\epsilon}\epsilon_R,V_{D}(y)+\sigma_{\epsilon}\epsilon_D\right\},  \label{eq:V_taste}
\end{equation}
where $\epsilon_i$ is i.i.d Extreme Value type I distributed with scale parameter $\sigma_{\epsilon}$. 

The government's value of repayment is given by
\begin{eqnarray}
V_{R}(b,y) & = & \max_{c\geq 0, j\in \mathcal{J},}\left\{u\left( c \right) + \sigma_{\varepsilon} \varepsilon_j + \beta
\mathbb{E}_{y^{\prime}|y}  V^{\sigma_{\epsilon}}(b_j^{\prime},y^{\prime }) \right\}, \label{eq:VR_taste} \\%
\text{subject to }&& \notag \\
c &=& y - (\delta+(1-\delta)z) b + q(b_j^{\prime},y)(b_j^{\prime} - (1-\delta)b).\notag
\end{eqnarray}
where $\varepsilon_j$ is i.i.d Extreme Value type I distributed with scale parameter $\sigma_{\varepsilon}$ and $V^{\sigma_{\epsilon}}$ is given by 
\begin{eqnarray}
V^{\sigma_{\epsilon}}(b,y)&=&E\left[\max\left\{ V_{R}(b,y)+\sigma_{\epsilon}\epsilon_R,V_{D}(y)+\sigma_{\epsilon}\epsilon_D\right\}\right] \notag \\
&=& \sigma_{\epsilon}\log\left\{\exp{\left(\frac{V_{R}(b,y)}{\sigma_{\epsilon}}\right)}+\exp{\left(\frac{V_{D}(b,y)}{\sigma_{\epsilon}}\right)}\right\}
,  \label{eq:EV_taste}
\end{eqnarray}

The value of default becomes 
\begin{eqnarray}
V_{D}(y) & = & u\left( c\right) + \beta \mathbb{E}_{y^{\prime}|y} \left[(1-\psi)V_{D}(y^{\prime}) + \psi V^{\sigma_{\epsilon}}(0,y^{\prime })\right], \ \ \label{eq:VD_taste} \\
\text{subject to } &&  \notag \\
 c & = & y - \phi(y). \notag 
\end{eqnarray}

The pricing function $q$ becomes
\begin{eqnarray}
q(b_i,y)  &=& \frac{\mathbb{E}_{y^{\prime}|y}\left[\left( 1 - d\left(b_i,y'\right)\right) \left[\delta + (1-\delta)\left(z+ \sum_j q\left( b_j,y^{\prime}\right)P(b_j|b_i,y^{\prime})\right)\right] \right]}{1+r}. \label{eq:q_taste}
\end{eqnarray}
where $P$ is the choice probability for next period debt conditional on the realization of endowment shock and, thanks to the particular assumption on the taste shocks, it is given by 
\begin{equation}    P(b_j|b_i,y)=\frac{\exp{\left(\tilde{V}_R(b_j,b_i,y)/\sigma_{\varepsilon}\right)}}{\sum_k\exp{\left(\tilde{V}_R(b_k,b_i,y)/\sigma_{\varepsilon}\right)}}
\end{equation}
where 
\begin{equation}
\tilde{V}_{R}(b_j,b_i,y)  =  u\left( y - (\delta+(1-\delta)z) b_i + q(b_j,y)(b_j - (1-\delta)b_i) \right) + \beta
\mathbb{E}_{y^{\prime}|y}  V^{\sigma_{\epsilon}}(b_j,y^{\prime })
\end{equation}

\section{Calibration \label{sec:calibration}}
We set the baseline parameters identical to that of \cite{Chatty} to make the model comparable to their printed results and we summarized them in Table \ref{Table_parameters}.

\begin{table}[ht] 
\centering
\begin{threeparttable}
\caption{Parameter Values\label{Table_parameters}}
\begin{tabular}{l c c}
\hline\hline
 & Parameter & Value \\   
\hline
Risk aversion   & $\gamma $ & 2 \\
Risk-free rate & $r$ & 1\% \\
Income autocorrelation coefficient & $\rho $ & 0.948503 \\
Standard deviation of innovations & $\sigma _{\epsilon }$ & 0.027092 \\
Mean log income & $\mu$ & (-1/2)$\sigma _{\epsilon }^{2}$ \\
Probability of reentry after default& $\psi$ & 0.0385 \\ %
Reciprocal of average maturity (CE) & $\delta $ &  0.05 \\
Coupon payments (CE) & $z$ & 0.03 \\
Coupon payments (HM) & $\kappa$ & $\frac{r+\delta}{1+r}$ \\
\hline
\multicolumn{3}{c}{Calibrated}\\
\hline
Discount factor & $\beta$ & 0.95402 \\
Income cost of defaulting  & $d_0$  & -0.18819 \\
Income cost of defaulting  & $d_1$  & 0.24558  \\
\hline\hline
\end{tabular}
\begin{tablenotes} [normal, flushleft]
\item \small 
\end{tablenotes}
\end{threeparttable}
\end{table}

\subsection{Numerical Solution}
This section briefly outlines the numerical algorithm undertaken in this paper to solve the model presented in the text while relegating the details to the Appendix. We solve the model using the value function iteration approach with global search methods, in particular using the BRENT method; value functions and price functions are iterated until the difference in two subsequent iterations remains the same.\footnote{We have also experimented using DUVMIF of IMSL which uses safeguarded interpolation to find the minimum of the objective function. As our results are identical, we are not reporting them here.} With this, a convergence criterion of $10^{-6}$ is attained with interpolation techniques or taste shocks. Yet, we fail to obtain convergence under the DSS, even with finer grids, 1,000 grids for assets and 1,000 grid points for income.\footnote{It is possible to obtain convergence with the DSS technique if one uses Tauchen's method to approximate the AR(1) process by setting the minimum and maximum income draws to be within 2.5 $\frac{\sigma_{\epsilon}}{1-\rho^2}$. Yet, unsurprisingly, this lowers the number of defaults in simulations.}

The solution of the model with taste shocks requires further elaboration. Although taste shocks are useful to smooth the kinks in the value and policy functions and obtain convergence in the solution of the model, they increase the computational time substantially. This is especially true when we introduce taste shocks for the debt choice. When taste shocks are introduced for the debt choice, conditional on the current debt choice and income, the debt policy becomes a density function as a function of the future debt choice as opposed to being a singleton in the absence of taste shocks. This not only increases the memory requirement for the policy function but also increases the computational time of the bond prices.\footnote{This issue arises in the presence of long-term debt. As can be seen in equation \ref{eq:q_taste}, with long-term debt, the bond price equation includes the resale value of the bond price which involves the density function of the bond choice. With one-period debt, the pricing equation will not involve that term.} \cite{Grey2019} proposes divide and conquer algorithm to speed up the computational time in a model with taste shocks. Although this algorithm speeds up the computation of the policy function, it does not help in the computation of the bond prices. In our applications, when the grid size is relatively high, the most computing intensive part of the code becomes the computation of the bond price.

\begin{figure}[H]
\begin{center}
\includegraphics[scale=.55]{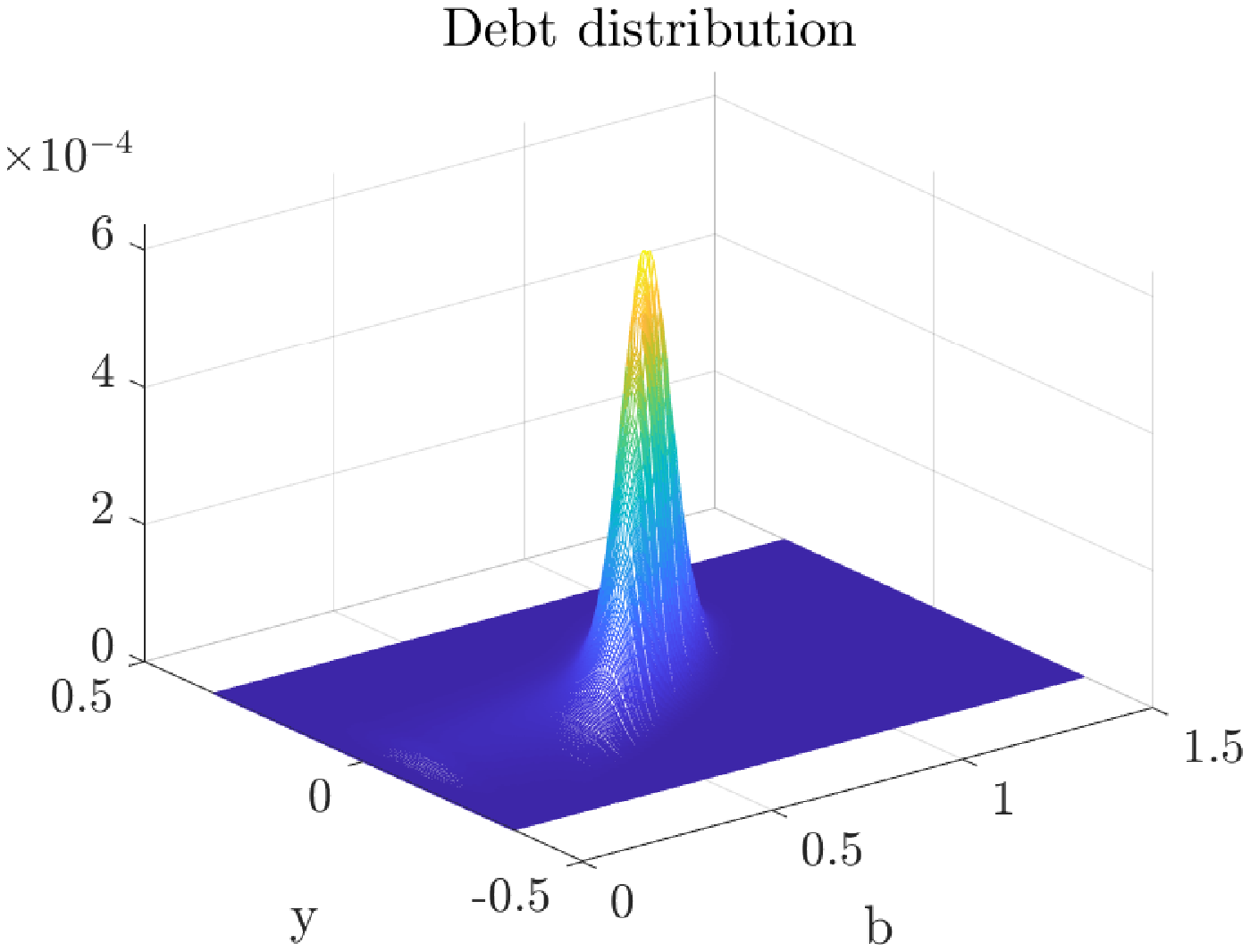}
\includegraphics[scale=.55]{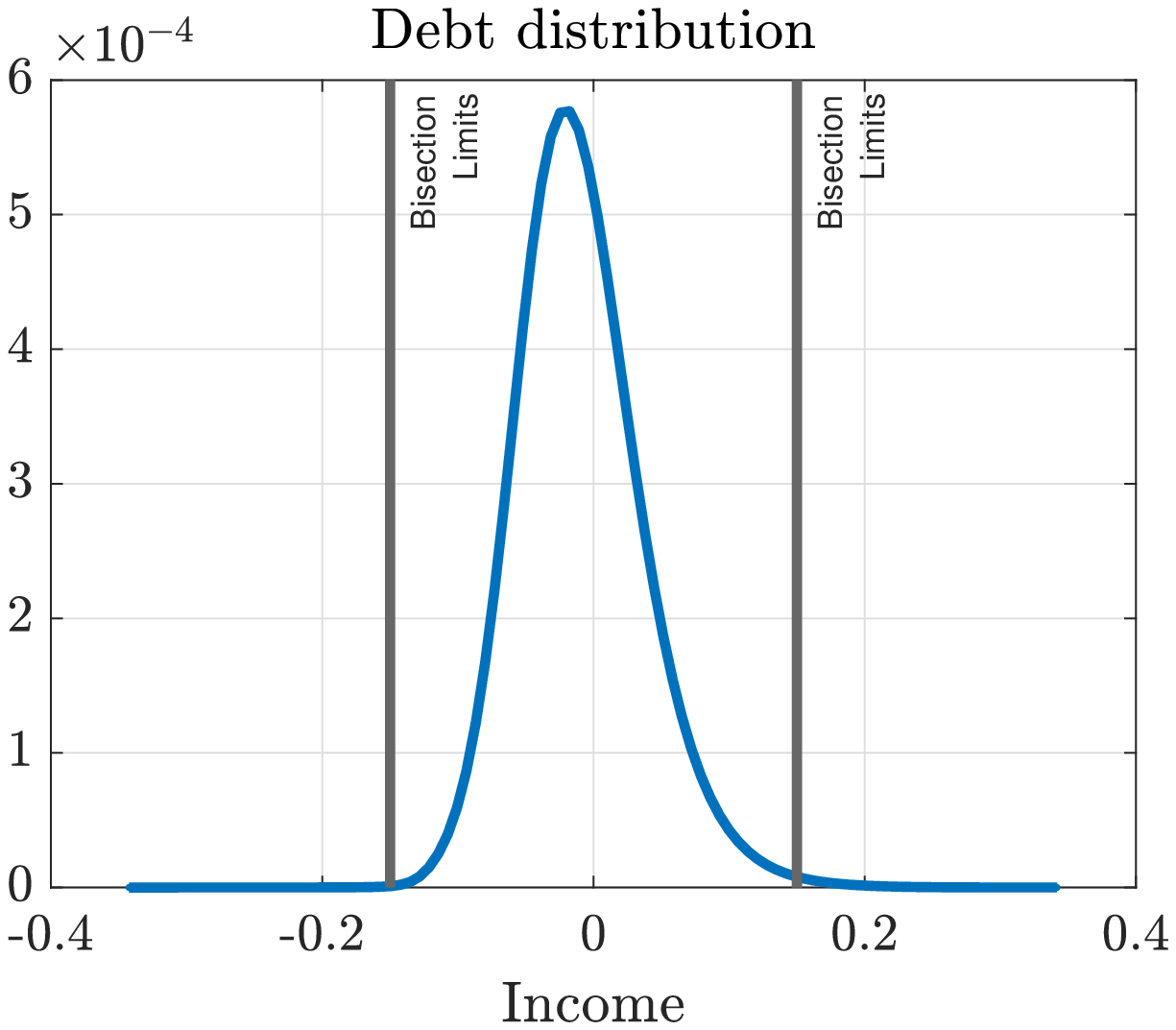}
\end{center}
\linespread{1.0} \caption{\small The left chart presents the non-contingent debt distribution and the right chart presents the debt distribution with debt level is set to its ergodic mean in the single asset problem. The embedded chart in the right shows how the the positive valued region is bisected. \label{fig:fdist}}
\end{figure}

We address these issues in two ways. First, we use the sparse matrices to store the bond policy functions. Second, we use the fact that for most bond choices, the choice probability will be computationally zero. In the computation of the value function, once we find out the grid point that attains the global maximum for the objective function, we only compute the choice probabilities for the adjacent grid points around the maximum.\footnote{The size of the adjacent grid points is set to $\min\{100 N_b \sigma_{\varepsilon},N_b/2-1\}$ where $N_b$ is the grid size for the debt choice and $\sigma_{\varepsilon}$ is the standard deviation of the taste shock.} As will be discussed below, the advantage of this method is greater in portfolio problems. To visualize the intuition behind this strategy, we provide Figure \ref{fig:fdist} which plots equilibrium debt distribution. The left chart plots it for the entire state space while the right chart bisects the left chart along the income axis for mean ergodic debt level observed in simulations. Notice in the left chart of the figure that debt distribution is only positive in a relatively smaller region while being entirely zero anywhere else. Thus, we do not need to store the entire choice probability but only the ones that are likely to be non-zero. The right chart visualizes how we implement this in a region where choice probability is non-zero. That is, we only make use of the positive valued region in the 2D graph using sparse matrices. Even though this speeds up the algorithm significantly and addresses the memory problems in the one-asset problem, it still falls short of delivering a reliable solution for the portfolio problem. 

\section{Quantitative results\label{sec:results}}
\begin{figure}[H]
\begin{center}
\includegraphics[scale=.45]{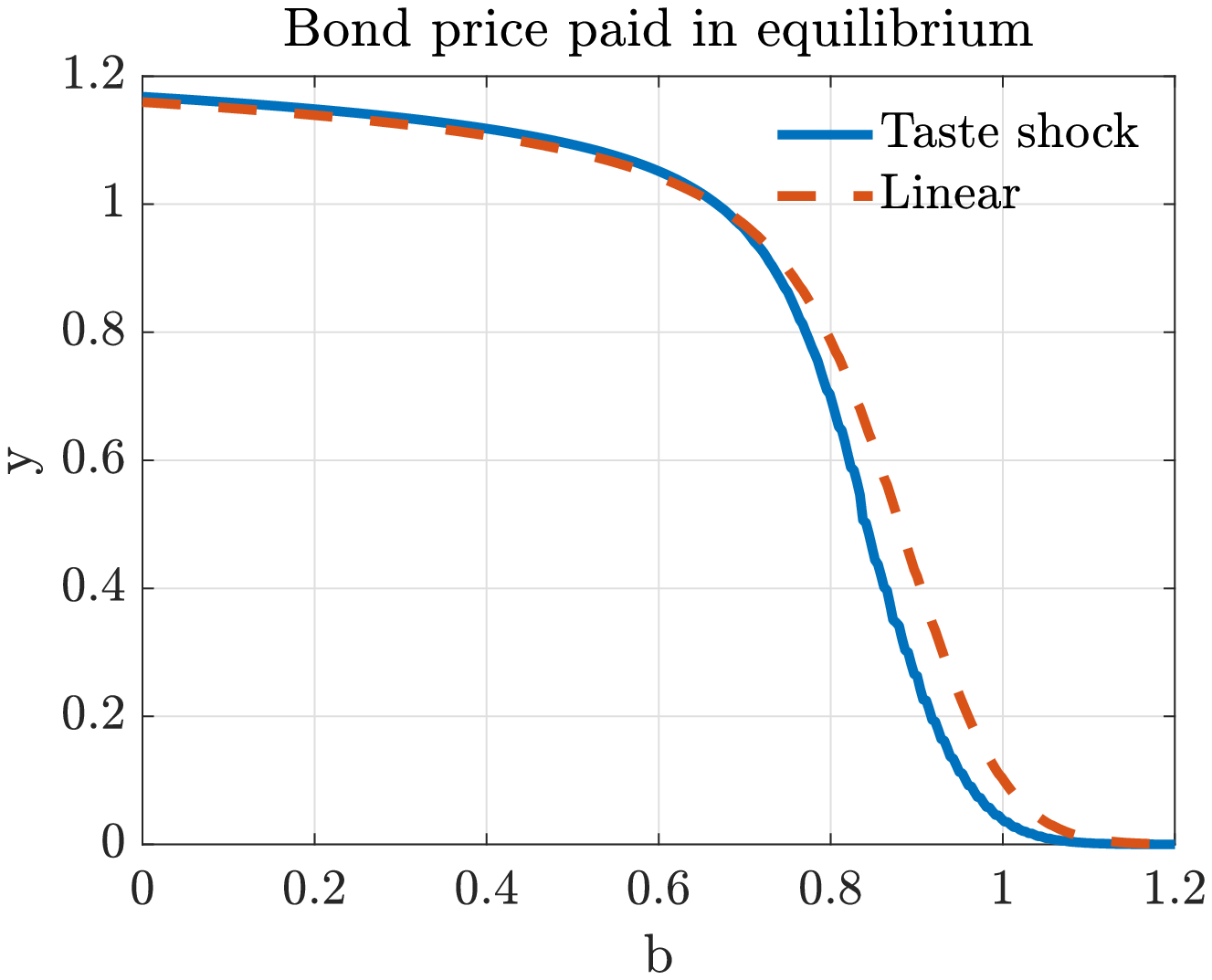}
\includegraphics[scale=.45]{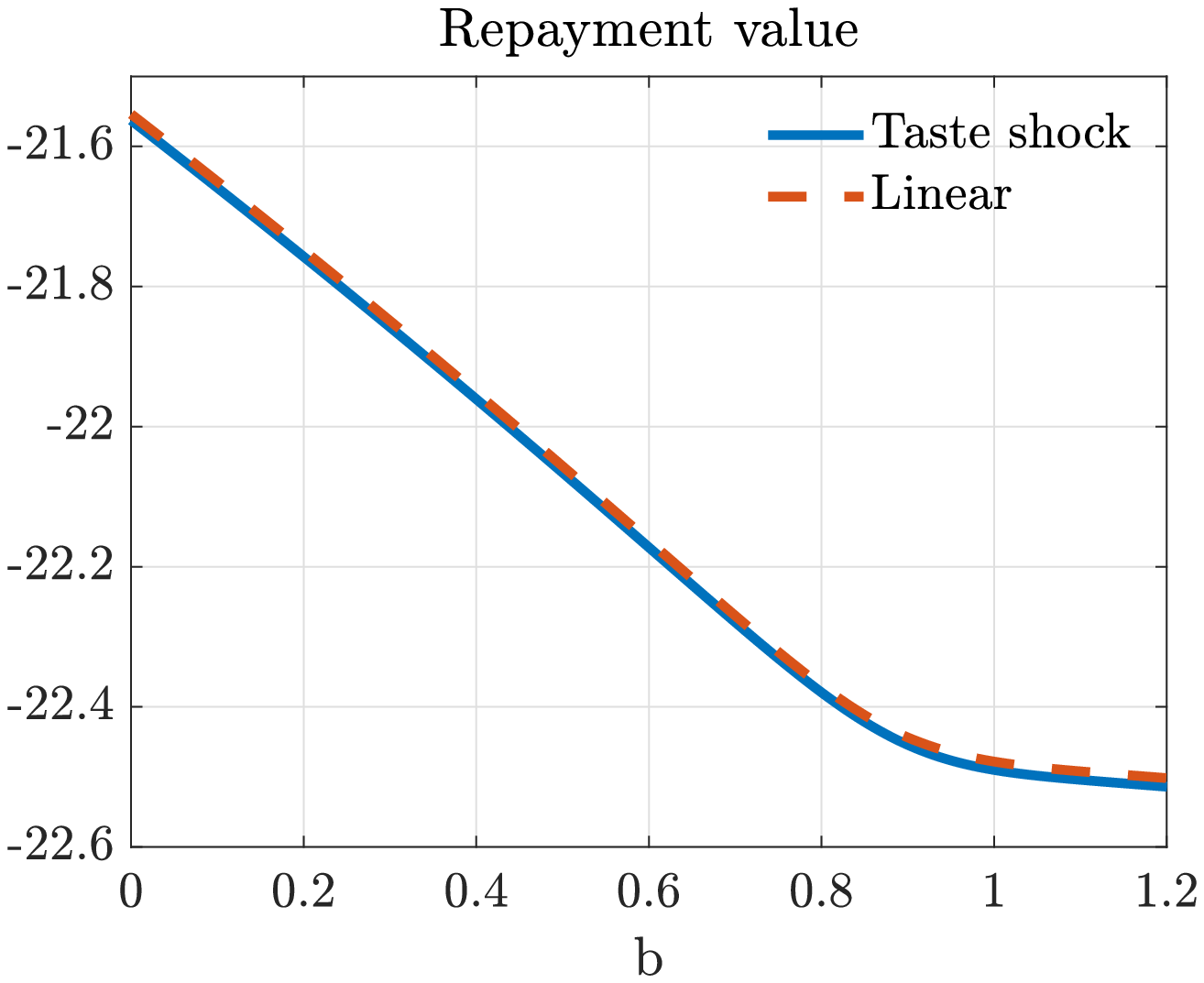}
\end{center}
\linespread{1.0} \caption{\small Policy functions in the single asset problem for mean income using linear interpolation and taste shocks. \label{fig:policy_1D}}
\end{figure}

\begin{figure}[H]
\begin{center}
\includegraphics[scale=.45]{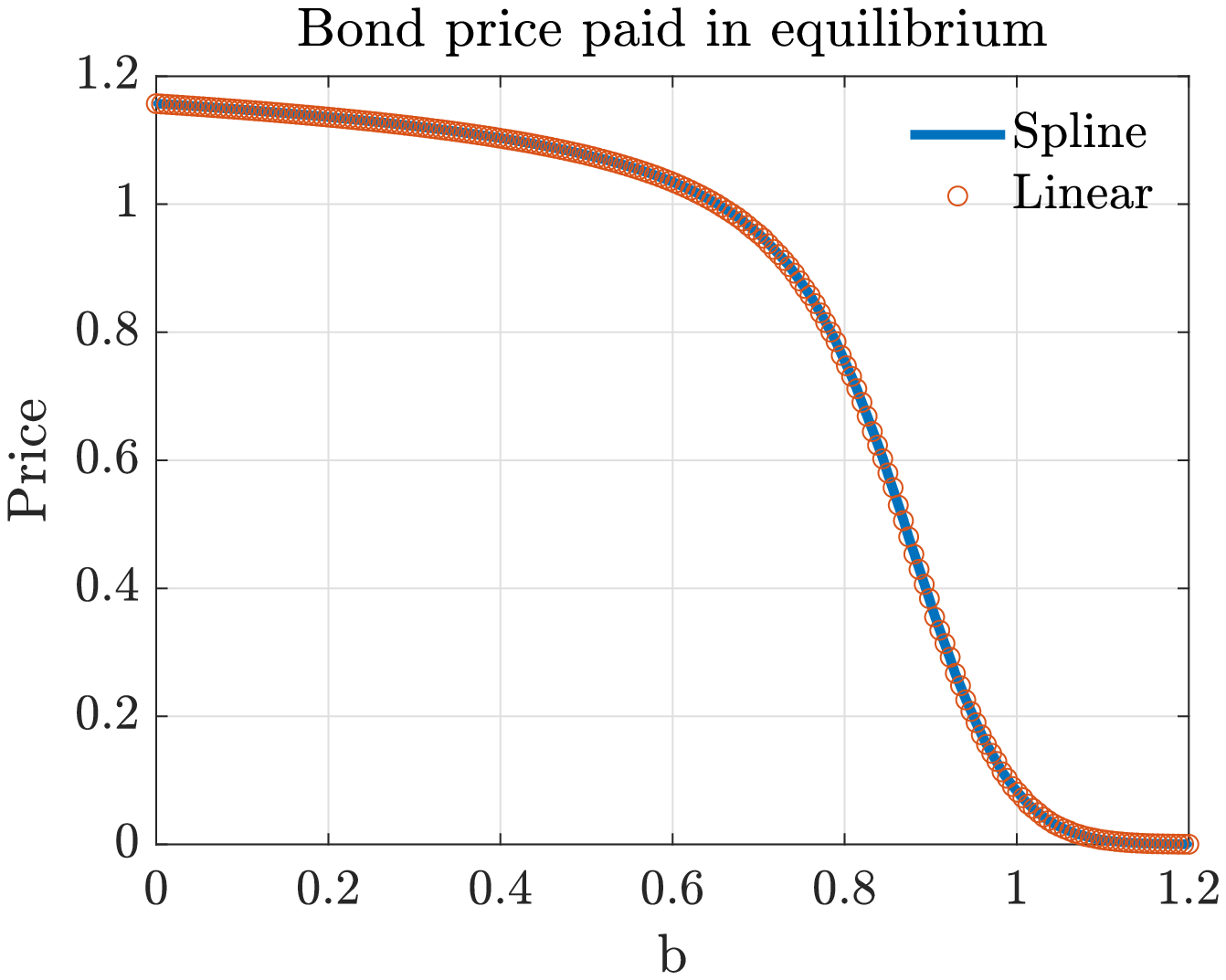}
\includegraphics[scale=.45]{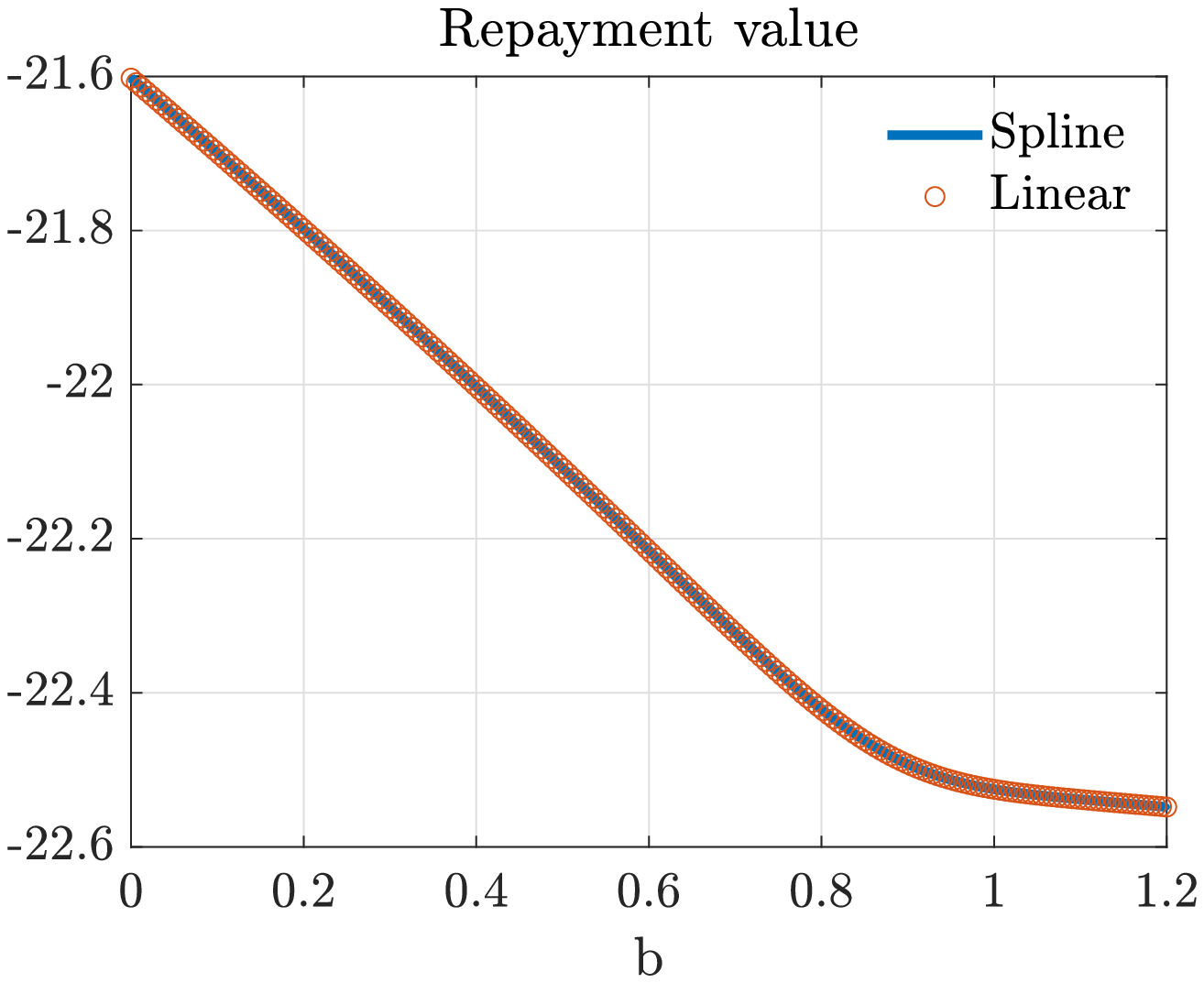}
\includegraphics[scale=.45]{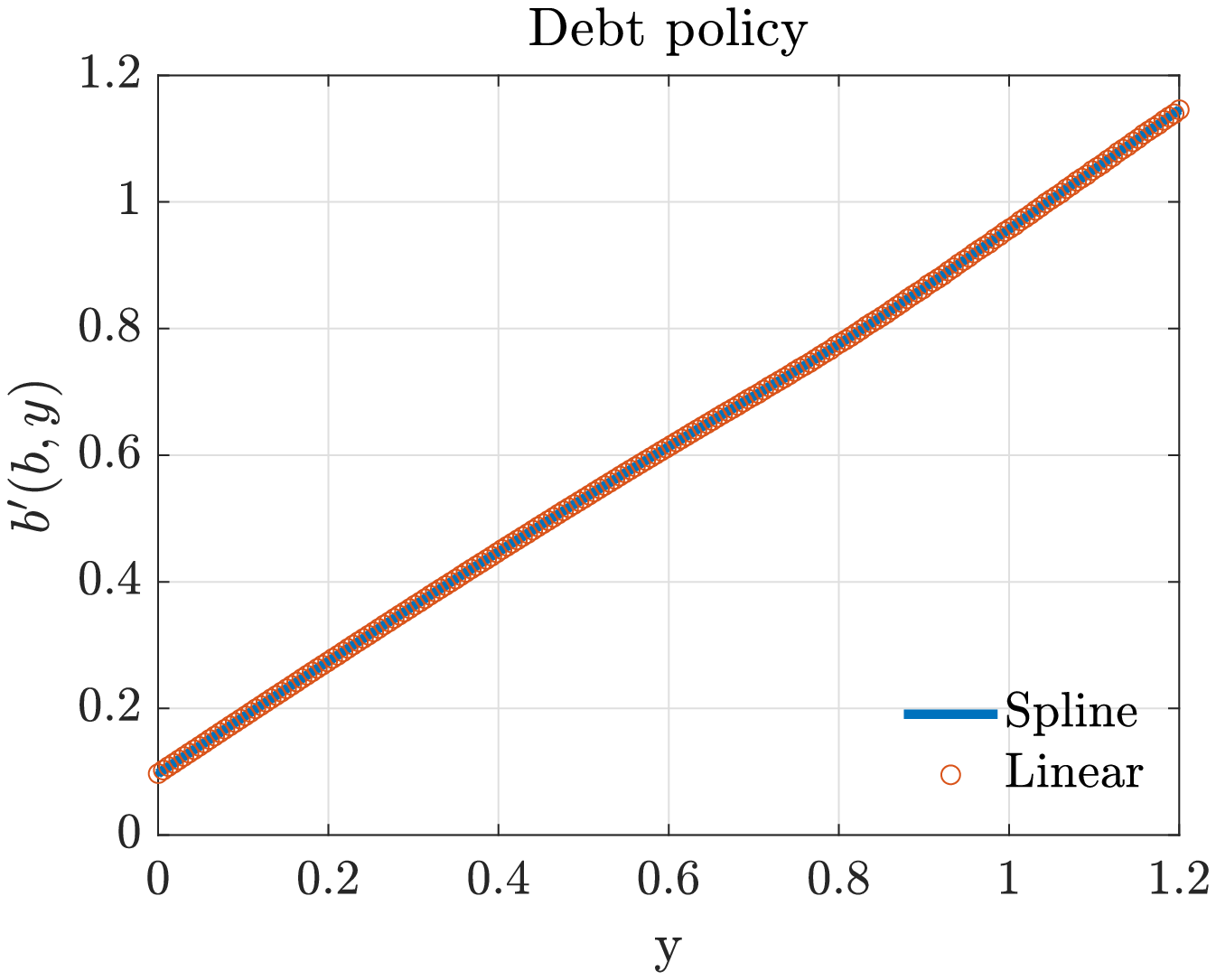}
\end{center}
\linespread{1.0} \caption{\small Policy functions in the single asset problem for mean income using linear and spline interpolations. \label{fig:policy_1D_ls}}
\end{figure}

Table \ref{tab:sim_results_CE} presents the results of alternative approximation techniques when using the formulation of CE. The first column reprints the results of CE that are in the publisher's website.\footnote{We also replicate it with the code which is available in the publisher's website.} The second column presents simulation results obtained with the DSS method without taste shocks. When we use an identical number of grid points that are used in CE, 200 for income shock and 350 for debt, our simulation results highly resemble CE even though the convergence cannot be obtained. In the third column, we introduce taste shocks and set the standard deviation of the taste shock to 0.003, which is the value used in CE. We can now obtain convergence but our moments are now somewhat changed. In particular, the debt-to-income ratio increases while the average spread and default rates decline. In columns (4) and (5), we solve the model using linear and B-spline interpolations with BRENT's minimization method. The moments under linear and spline interpolations are highly similar while being different from the DSS, particularly default and spread moments. We lastly plot the policy functions, in particular, price and value functions in Figures \ref{fig:policy_1D} and \ref{fig:policy_1D_ls}. Figure \ref{fig:policy_1D} plots policy functions using taste shocks and linear interpolation, whereas Figure \ref{fig:policy_1D_ls} plots policy functions obtained with linear and spline interpolations. Notice that Figure \ref{fig:policy_1D} does not include the debt policy function as debt policy becomes a distribution under taste shocks. Even though simulation moments are very similar for both taste shocks linear interpolation, there is still visually visible differences in policy functions. Yet, this difference is not visible between linear and spline interpolations.

To better understand the role of taste shocks, as this is often a technique being used in quantitative literature to boost the convergence properties of the algorithm and is also the technique being used in CE, we undertake a number of analyses. In particular, we rerun our analysis by increasing the grid sizes and reducing the standard deviation of the taste shock using our own algorithm as the CE code provided in the publisher's website runs into memory problem when grid sizes are expanded. Results of this analysis are presented in Table \ref{tab:sim_results_taste}. To start with, we first reduce the standard deviation of the taste shock from 0.003 to $10^{-4}$ in column (2) while keeping grid sizes in the baseline scenario identical. Our results are now very similar to the printed results of CE. Next, we increase the grid sizes, in particular, we set the grid numbers for assets to 1000 and for income to 500 in subsequent columns. In the third column, we solve our model with the DSS and convergence again fails. We then introduce taste shocks in columns (4) and (5) with a standard deviation of taste shocks of $10^{-4}$ and $5\times10^{-5}$, respectively. The model does not converge when we set it to $10^{-5}$. It appears that except the default frequency, the model with taste shocks whose value is small enough just to obtain convergence may generate similar results to that of linear and spline approximations. Yet, with DSS and taste shocks, obtaining consistent and reliable results is still a challenge because it appears that every time we increase grid sizes, moments still move, albeit small, and then the researcher needs to experiment a number of different values of taste shocks to find the smallest value that would provide convergence. With linear or spline interpolations, this is not the case. Changing the grid size does not move the model moments for high enough grid points. Even though both interpolation techniques yield highly similar results, linear interpolation is at least 20 times faster than B-splines.

Now it remains to present the main difference between the CE and the HM environments. HM uses Macaulay duration which is defined as the weighted average maturity of future cash flows and formally provided in footnote \ref{footnote_spread}, whereas CE defines it as the reciprocal of probability of debt maturing next period, which is $\frac{1}{\delta}$. This leads to differences in how the internal rate of return or yield-to-Maturity is computed in two environments. CE computes it such that the internal rate of return $r(b^{\prime},y)$ satisfies $q(b^{\prime}) = \frac{\delta+(1-\delta)z}{\delta+r(b^{\prime},y)}$. The annualized spread then can be provided as $(1+r(b^{\prime},y))^4 - (1+r)^4$. The yield and the spread in the HM world is again provided in footnote \ref{footnote_spread}. To make both frameworks comparable, we set $z$, which is the coupon payment in the CE world, to $\frac{r}{1+r}$, and keep everything else in Table \ref{Table_parameters} constant. With this transformation, both environments become identical and results are presented in Table \ref{tab:sim_results_HM}. Differences in both tables are attributed to how duration and thus spreads are computed. As noted before, HM divides debt-to-income ratio by 4 in a quarterly model while CE does not.

\section{Portfolio problem}

We introduce a new asset to the model presented in the text to be able to compare solutions methods under a portfolio allocation problem. This is an important and intentional extension to highlight two main points: $(i)$ the current state of the literature now often requires to solve for more than one asset, and $(ii)$ the solution method used matters. In particular, we now allow the sovereign to borrow a one-period non-defaultable asset which has a risk-free price. To formalize let $a$ denote the amount of non-defaultable bonds the sovereign is obliged to pay and $\bar{a}$ denote the upper limit the sovereign can issue. We set it to be 10 percent of the mean trend income. We drop the superscripts and use the CE formulation as our preferred approach. With this simplistic portfolio problem, we can write the recursive formulation as follows:
\begin{equation}
V(b,a,y)=\underset{}{Max }\left\{ V_{R}(b,a,y),V_{D}(a,y)\right\},  \label{eq:V}
\end{equation}
where the government's value of repayment is given by
\begin{eqnarray}
& & V_{R}(b,a,y)  =  \max_{c\geq0, b^{\prime}, a^{\prime}}\left\{u\left( c \right) +  \beta
\mathbb{E}_{y^{\prime}|y}  V(b^{\prime},a^{\prime},y^{\prime }) \right\}, \label{eq:VR} \\%
&&\text{subject to } \notag \\
&& c = y - (\delta+(1-\delta)z) b + q(b^{\prime},a^{\prime},y)(b^{\prime} - (1-\delta)b)-a+\frac{a^{\prime}}{1+r},\\ \notag
&&a^{\prime} \leq \bar{a}.\notag
\end{eqnarray}
As these bonds are non-defaultable, in the event of a default, the sovereign is obliged to repay in full. Thus, the value of default reads
\begin{eqnarray}
&& V_{D}(a, y)  =  u\left( c\right) + \beta \mathbb{E}_{y^{\prime}|y} \left[(1-\psi)V_{D}(0,y^{\prime}) + \psi V(0,0,y^{\prime })\right], \ \ \label{eq:VD} \\
&& \text{subject to } \notag \\
&& c = y - a- \phi(y). \notag 
\end{eqnarray}
The solution to this problem yields a binary default decision rule $\hat{d}(b,a,y)$ $\in$ $\{0,1\}$, and borrowing rules $\hat{b}(b,a,y)$, $\hat{a}(b,a,y)$. In equilibrium, lenders use these decision rules to price contracts. Price functions for non-contingent bonds $q$ solve the following functional equation:
%\fontsize{10}{15}{
\begin{eqnarray}
q(b^{\prime},a^{\prime},y)  &=& \frac{\mathbb{E}_{y^{\prime}|y}\left[\left( 1 - \hat{d}\left(b^{\prime},a^{\prime},y'\right)\right) \left[\delta + (1-\delta)\left(z+ q\left( b^{\prime\prime},a^{\prime\prime},y^{\prime}\right)\right)\right] \right]}{1+r}. \label{eq:q}
\end{eqnarray}
\subsection{Solution Algorithm}

When using interpolation techniques, a built-in minimization routine should be utilized instead of taste shocks. This is particularly essential when solving for the optimal portfolio problem. Among the multidimensional minimization methods to solve a constrained optimization problem, we mainly experimented with four common routines: the Downhill simplex method due to Nelder and Mead, Powell's method with direction-set methods, the quasi-Newton Davidon-Fletcher-Powell (DFPMIN) algorithm, and the BOBYQA routine (\cite{BOBYQA}). Except for the DFPMIN routine, these are the methods of choice when researchers cannot easily compute derivatives, as in our application. We also experimented with the Broyden-Fletcher-Goldfarb-Shanno (BFGS) algorithm, which is closely related to DFPMIN, and provided approximations of partial derivatives by finite differences.

Both B-spline and linear interpolations fail to deliver convergence without these routines, so their relative performance becomes important. Following important observations stand out. The Powell routine complemented with B-splines is by far the most superior alternative, while BOBYQA is a close runner-up. In some instances, one algorithm becomes superior to the other depending on the calibration and grid size. Although we find Powell to be more viable in attaining convergence and it is our preferred method, we observed instances where Powell routine explodes miserably during a local search, which is invoked if the algorithm fails to find a global optimum within a certain number of iterations. During a local search, we bracket the solution to be lying within a proximity of candidate optima obtained with global solution method which is outlined in Appendix. Thus, the code fails to converge for some particular parameter sets and grid sizes. This occurs because the Powell routine does not require boundaries as inputs, but rather directions. If it fails to find the local maxima in a given direction, it explodes. Although researchers can include a penalty in their user-supplied objective function to penalize the routine if it goes out of bounds, this does not prevent the routine from experimenting outside of boundaries. BOBYQA, however, requires boundaries as inputs, ensuring it stays within defined search space. However, it may lag behind the Powell algorithm in attaining convergence. Despite this, both of these routines are strong and can be utilized as near-perfect substitutes since they generate near-identical results.

In general, the algorithm is more likely to converge with splines than with linear interpolation, although this comes at a higher computational cost. If our algorithm with linear interpolation fails to converge within the usual 250 iterations, we switch to spline interpolation using the obtained value and policy functions with linear interpolation as initial guesses for the model with splines.

When using spline interpolation, it's worth noting that we undertake linear interpolation over income levels and spline interpolation over defaultable and non-defaultable debt levels. This is because three-dimensional spline interpolations are significantly more time-consuming. As such, we use two-dimensional B-splines. We're happy to provide codes upon request for three-dimensional B-splines.

Other algorithm and the quasi-Newton Davidon-Fletcher-Powell, both failed to deliver convergence by a big margin. Thus, we are not presenting these results.

With taste shocks, we also fail to obtain convergence and fail having systematically consistent results. That is, as we increase the grid space, up until facing memory problems, our simulations results keep constantly changing. Thus, our simulation results from the introduction of taste shocks are not provided as they are not reliable. 

\subsection{Results of the portfolio problem}
Table \ref{tab:sim_results_CE} presents the simulation results. Note that we kept the parameter space of the CE world unchanged. Leaving the normative implications of introducing non-defaultable debt aside (see \cite{Onderswap}), we solve this model with linear and spline interpolations using Powell and Bobyqa minimization routines. 

\begin{figure}[H]
\begin{center}
\includegraphics[scale=.45]{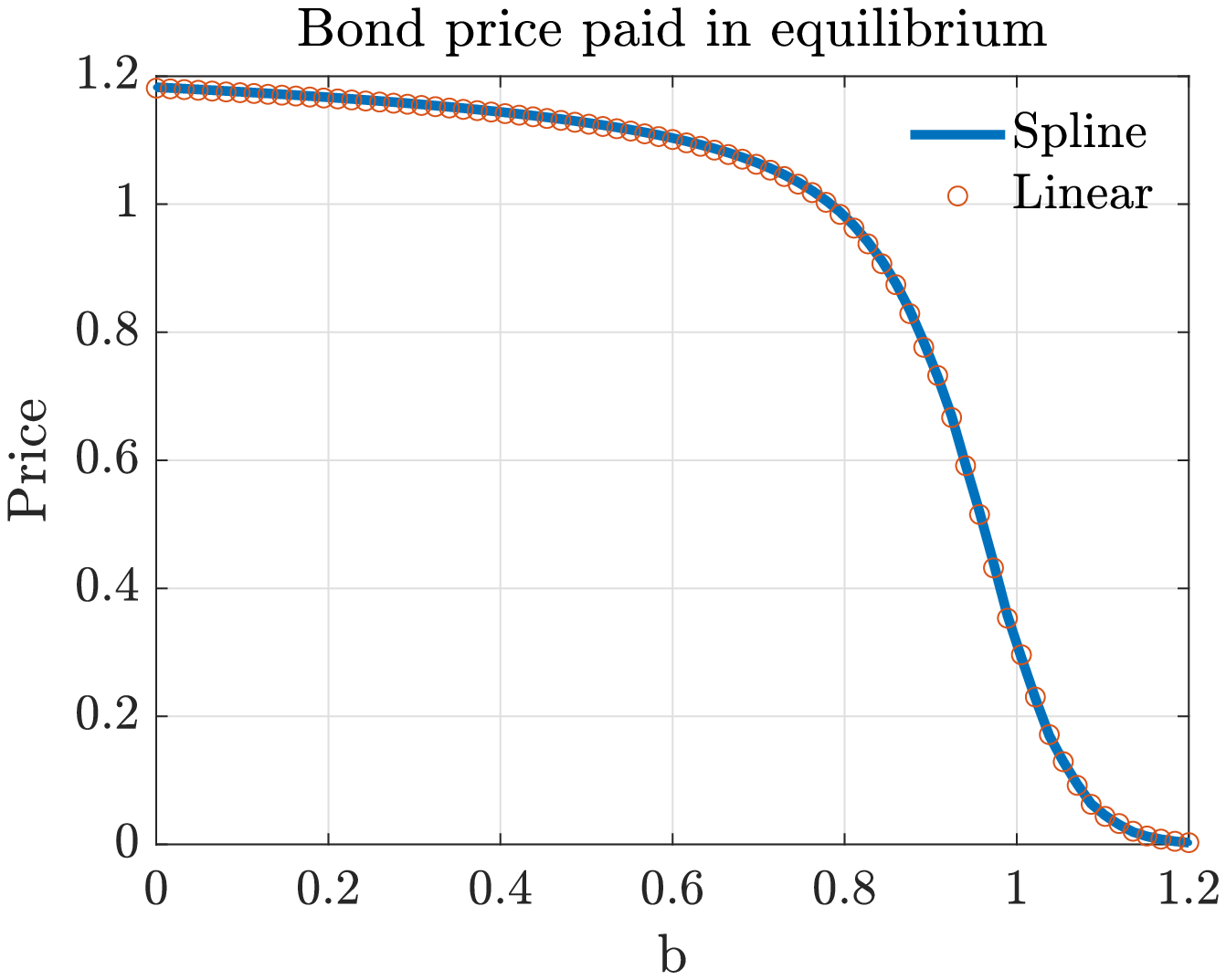}
\includegraphics[scale=.45]{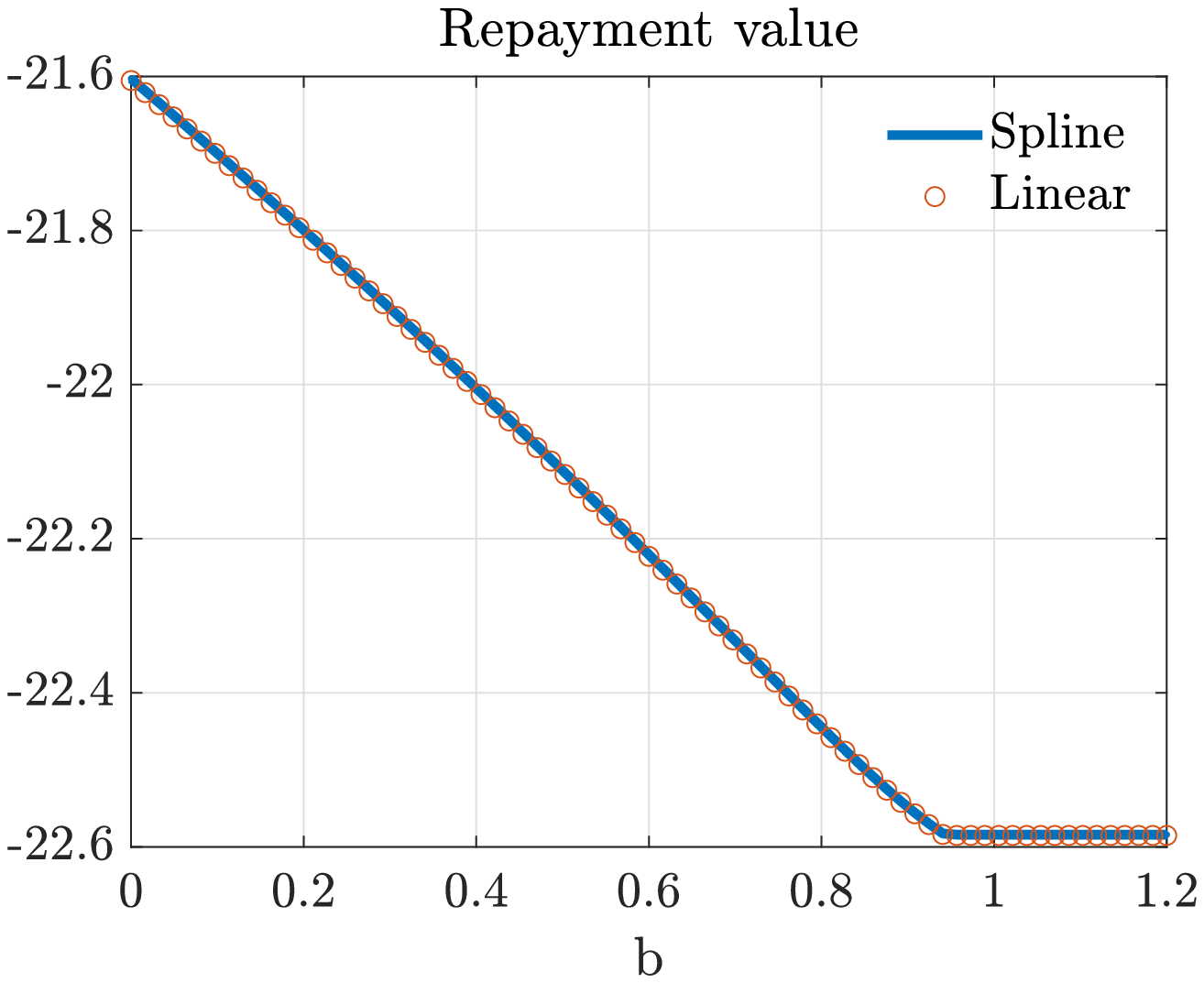}
\includegraphics[scale=.45]{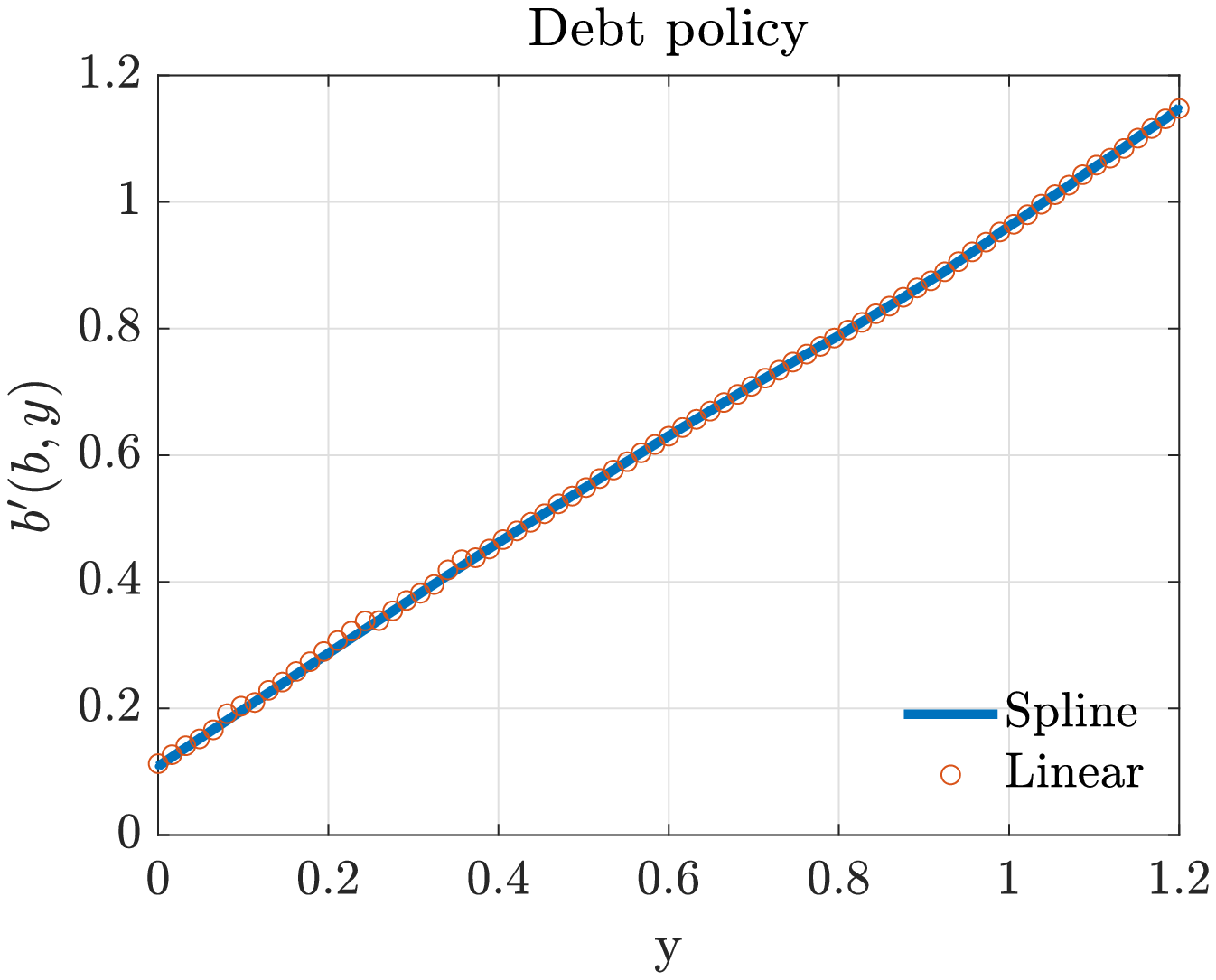}
\end{center}
\linespread{1.0} \caption{\small Policy functions in the portfolio problem. All are obtained with the Powell algorithm. The figures are plotted for mean income and the initial stock of non-defaultable debt is set to the ergodic level of non-defaultable debt stock. \label{fig:policy_2D}}
\end{figure}

Both linear and spline interpolation techniques as well as multidimensional minimization routines yield highly similar moments in Table \ref{tab:sim_results_CE}. We also plot the policy functions in Figure \ref{fig:policy_2D}. The figures are plotted for mean income and the initial stock of non-defaultable debt is set to the ergodic level of non-defaultable debt stock. The figure shows that price, value and default debt policy functions are almost idential. The linear interpolation takes around 49 hours with OPenMP.

\section{Conclusion\label{sec:conclusion}}
We evaluate and extend on the solution methods of models with binary and continuous choice variables in dynamic programming problems, particularly problems where discrete state space solution methods are not viable. For this, we apply these methods in long-term debt and default. We extend on the implementation of taste shocks and compare performance of multidimensional black-box optimizers,  that are used by researchers in applied mathematics literature to benchmark their algorithm's performance, along with linear and spline interpolation. We mainly compare Powell and BOBYQA algorithms. We show that solution method matters. Taste shocks fail attaining convergence in multidimensional problems and often run into memory problems. Linear interpolation techniques are 20 times faster than high dimensional spline interpolation but sometimes lags behind to become a viable option on attaining convergence. Thus, we recommend researchers to initiate their methods with a linear interpolation and switch to spline if linear interpolation does not converge within certain number of iterations. Among multidimensional algorithms, BOBYQA lags behind Powell routine in terms of viability in attaining global maximum while we have always failed attaining convergence with Nelder-Mead downhill simplex algorithm and quasi-Newton Davidon-Fletcher-Powell (DFPMIN) algorithm.

%\begin{landscape}
\begin{table}[htbp!]
\centering
%\begin{adjustbox}{max width=5in,totalheight=9in}
%\resizebox{\linewidth}{!}{% 
\begin{threeparttable}
\caption{Simulation results with CE formulation\label{tab:sim_results_CE}}
\begin{tabular}{l*{6}{c}}
\hline\hline
    %&\multicolumn{4}{c}{CE}&\multicolumn{3}{c}{HM}\\
		%\cline{1-2}
		%\cline{5-7}
&\multicolumn{1}{c}{(1)}&\multicolumn{1}{c}{(2)}&\multicolumn{1}{c}{(3)}&\multicolumn{1}{c}{(4)}&\multicolumn{1}{c}{(5)}\\

& CE	& DSS 	& Taste shock  & Linear & B-Spline \\   
\hline
\multicolumn{6}{c}{Debt Statistics}\\
\hline
Debt-to-Income   ratio             & 69.98 & 70.84 & 73.9  & 70.69 & 70.57 \\
Average spread                     & 8.15  & 7.94  & 7.46  & 8.39  & 8.47  \\
SD of spread          & 4.43  & 2.49  & 2.91  & 3.47  & 3.5   \\
Default rate          & 6.75  & 6.68  & 5.06  & 5.83  & 5.79  \\
%Duration										& 4     & 3.97  & 3.92   & \\
\hline
\multicolumn{6}{c}{Business Cycle Statistics}\\
\hline
$\sigma(c)/\sigma(y)$	 & 1.11  & 1.31  & 1.32  & 1.32  & 1.32  \\
$\sigma(tb/y)/\sigma(y)$ & 0.17  & 0.35  & 0.36  & 0.34  & 0.34  \\
corr(c,y)          		& 0.98  & 0.99  & 0.99  & 0.99  & 0.99  \\
corr(tb/y,y)			& -0.88 & -0.83 & -0.83 & -0.87 & -0.87 \\
corr(spread,y)			& -0.79 & -0.81 & -0.75 & -0.75 & -0.75\\
\hline\hline
\end{tabular}
\begin{tablenotes} [normal, flushleft]
\item \small The first column reprints the results of CE. Column (2) is obtained using the DSS approximation while column (3) is obtained adding taste shock to the DSS approximation. Columns (4) and (5) are obtained with linear and B-spline interpolations, respectively. Convergence cannot be attained for the DSS column. All column moments are obtained using CE's long-term debt formulation and the details about the simulations are relegated to the Appendix.
\end{tablenotes}
\end{threeparttable}
%\end{adjustbox}
%}
\end{table}
%\end{landscape}

\begin{table}[htbp!]
\centering
%\begin{adjustbox}{max width=5in,totalheight=9in}
%\resizebox{\linewidth}{!}{% 
\begin{threeparttable}
\caption{Simulation results with CE formulation\label{tab:sim_results_taste}}
\begin{tabular}{l*{6}{c}}
\hline\hline
    %&\multicolumn{4}{c}{CE}&\multicolumn{3}{c}{HM}\\
		%\cline{1-2}
		%\cline{5-7}
&\multicolumn{1}{c}{(1)}&\multicolumn{1}{c}{(2)}&\multicolumn{1}{c}{(3)}&\multicolumn{1}{c}{(4)}&\multicolumn{1}{c}{(5)}\\
& CE	& $\sigma_t = 10^{-4}$  & DSS  & $\sigma_t = 10^{-4}$ & $\sigma_t = 5\times10^{-5}$  \\
& CE	& (350-200) & (1000-500)  & (1000-500) &  (1000-500) \\
   
\hline
\multicolumn{6}{c}{Debt Statistics}\\
\hline
Debt-to-Income   ratio   & 69.98 & 69.98 & 71.12 & 70.71 & 70.67 \\
Average spread           & 8.15  & 8.18  & 8.46  & 8.31  & 8.32  \\
SD of spread          & 4.43  & 3.34  & 3.52  & 3.41  & 3.42  \\
Default rate          & 6.75  & 6.65  & 6.58  & 6.62  & 6.65  \\
%Duration										& 4     & 3.97  & 3.92   & \\
\hline
\multicolumn{6}{c}{Business Cycle Statistics}\\
\hline
$\sigma(c)/\sigma(y)$	 & 1.11  & 1.32  & 1.33  & 1.32  & 1.32  \\
$\sigma(tb/y)/\sigma(y)$& 0.17  & 0.36  & 0.36  & 0.35  & 0.35  \\
corr(c,y)          		& 0.98  & 0.99  & 0.99  & 0.99  & 0.99  \\
corr(tb/y,y)			& -0.88 & -0.83 & -0.88 & -0.86 & -0.86 \\
corr(spread,y)			& -0.79 & -0.75 & -0.76 & -0.75 & -0.75 \\
\hline\hline
\end{tabular}
\begin{tablenotes} [normal, flushleft]
    \item \small The first column reprints the results of CE. Column (2) is obtained using taste shocks with the standard deviation of the taste shock value of $10^{-4}$ using 350 grid points on debt and 200 grid points on income. Column (3) is obtained with DSS approximation using 1000 grid points on debt and 500 grid points on income while column (4) is obtained adding taste shock to this DSS approximation with a value of $10^{-4}$. Columns (5) is obtained with the standard deviation of the taste shock value of $5\times10^{-5}$. Convergence cannot be attained for the DSS column. All column moments are obtained using CE's long-term debt formulation and the details about the simulations are relegated to the Appendix.
\end{tablenotes}
\end{threeparttable}
%\end{adjustbox}
%}
\end{table}

\begin{table}[htbp!]
\centering
%\begin{adjustbox}{max width=5in,totalheight=9in}
%\resizebox{\linewidth}{!}{% 
\begin{threeparttable}
\caption{Differences between the HM and the CE formulation\label{tab:sim_results_HM}}
\begin{tabular}{l*{9}{c}}
\hline\hline
&\multicolumn{1}{c}{(1)}&\multicolumn{1}{c}{(2)}\\
& CE linear	& HM linear \\ 
\hline
\multicolumn{3}{c}{Debt Statistics}\\
\hline
Debt-to-Income   ratio & 93.34 & 93.35 \\
Average   spread      & 8.44  & 8.11  \\
SD of spread          & 3.49  & 3.35  \\
Default   rate        & 5.8   & 5.8   \\
Duration (quarters)   & 20    & 13 \\
\hline
\multicolumn{3}{c}{Business Cycle Statistics}\\
\hline
$\sigma(c)/\sigma(y)$ & 1.32  & 1.32  \\
$\sigma(tb/y)/\sigma(y)$ & 0.34  & 0.34  \\
corr(c,y)   & 0.99  & 0.99  \\
corr(tb/y,y)& -0.87 & -0.87 \\
corr(spread,y)	& -0.75 & -0.75 \\
\hline\hline
\end{tabular}
\begin{tablenotes} [normal, flushleft]
\item \small The first column reports the results of CE and the second column reports the results of HM. Both columns are obtained using linear interpolation. 
\end{tablenotes}
\end{threeparttable}
%\end{adjustbox}
%}
\end{table}

\begin{table}[htbp!]
\centering
%\begin{adjustbox}{max width=5in,totalheight=9in}
%\resizebox{\linewidth}{!}{% 
\begin{threeparttable}
\caption{Simulation results with portfolio allocation\label{tab:sim_results_portfolio}}
\begin{tabular}{l*{9}{c}}
\hline\hline
&\multicolumn{2}{c}{Powell}&\multicolumn{2}{c}{Bobyqa}\\
&\multicolumn{1}{c}{(1)}&\multicolumn{1}{c}{(2)}&\multicolumn{1}{c}{(3)}&\multicolumn{1}{c}{(4)}\\
& Linear	& B-spline & Linear	& B-spline \\ 
\hline
\multicolumn{5}{c}{Debt Statistics}\\
\hline
Debt-to-Income   ratio & 78.10  & 78.22  & 78.12& 78.25\\
Non-defaultable-debt-to-Income   ratio & 9.68 & 9.68 & 9.67 & 9.67   \\
Average   spread      & 7.22  & 7.13  & 7.12& 7.08 \\
SD of spread          & 2.80  & 2.78 & 2.75 & 2.76  \\
Default   rate        & 4.43    &4.44  & 4.37   &4.44  \\
\hline
\multicolumn{5}{c}{Business Cycle Statistics}\\
\hline
$\sigma(c)/\sigma(y)$ & 1.34  & 1.34 & 1.34 & 1.34  \\
$\sigma(tb/y)/\sigma(y)$ & 0.37  & 0.37 & 0.37 & 0.37 \\
corr(c,y)   & 0.99  & 0.99 &  0.99 &  0.99  \\
corr(tb/y,y)& -0.86 &  -0.86 & -0.86 & -0.86  \\
corr(spread,y)	& -0.75 & -0.75 &  -0.75&  -0.75 \\
\hline\hline
\end{tabular}
\begin{tablenotes} [normal, flushleft]
\item \small The first and the second column moments are obtained using Powell minimization routine while the third and and the fourth column moments are obtained with Bobyqa. 
\end{tablenotes}
\end{threeparttable}
%\end{adjustbox}
%}
\end{table}

\newpage
\newpage
\linespread{1.0}
\bibliographystyle{aea} %Style. Alternative: chicago. Or try \bibliographystyle{elsart-harv} for the Elsevier journal bibliography style.
\bibliography{bib_default}
%\newpage
\appendix 
\setcounter{table}{0}
\setcounter{figure}{0}
\renewcommand{\thetable}{A\arabic{table}}
\renewcommand{\thefigure}{A\arabic{figure}}
\section{Appendix}
\linespread{1.0}
\subsection{Numerical algorithm\label{sec:computation}}
The computational algorithm used in this paper requires iterating on the value and price functions until a convergence criteria of $10^{-6}$ is obtained. To evaluate the functions, equally spaced grid points; to approximate the functions outside the grid points linear or spline interpolation is used. To evaluate the expectations 200 Gauss-Legendre quadrature points are utilized. Below, we first present our algorithm for the one-asset model and then provide the details of the portfolio allocation problem.
\subsubsection{One-asset model}
\begin{enumerate}
\item \textbf{Initialization:} Initial guesses of $V_R$, $V_D$, $q$  are set as the last period of finite-horizon economy as follows: $V_R(b,y) = u(y - \kappa b)$, $v_D(y) = u(y - \phi^d(y))$, and $q = 0$.
\item \textbf{Global stage:} Optimization problem defined in the equation (\ref{eq:VR_CE}) is solved for each grid point on assets and income and then I search for a globally optimum point for next period's borrowing decisions for non-contingent debt. This requires generating 100 grid points for $b^{\prime}$ (and $a^{\prime}$ in the portfolio problem) to find the maximizing candidates. We feed that particular grid point into one-dimensional BRENT routine as an initial guess to pinpoint the optimal $b^{\prime}$ with a double precision. We have also experimented with a safeguarded quadratic interpolation method DUVMIF of IMSL and obtained identical results.
\item \textbf{Stopping rule:} Iterate the procedure defined above for equations (\ref{eq:VR_CE}) to (\ref{eq:q_CE}) until convergence criteria is met. 
\end{enumerate}

\subsubsection{Portfolio model}
\begin{enumerate}
\item \textbf{Initialization:} Initial guesses of $V_R$, $V_D$, $q$  are set as the last period of finite-horizon economy as follows: $V_R(b,a,y) = u(y - \kappa b -a)$, $v_D(y) = u(y - \phi^d(y) - a)$, and $q = 0$.
\item \textbf{Global stage:} Optimization problem defined in the equation (\ref{eq:VR}) is solved for each grid point on assets and income and then I search for a globally optimum point for next period's borrowing decisions for non-contingent debt. This requires generating 100 grid points for $b^{\prime}$ and $a^{\prime}$ to find the maximizing candidates. We feed that particular grid points into one-dimensional BRENT routine as an initial guess to pinpoint the optimal $b^{\prime}$ with a double precision. We have also experimented with a safeguarded quadratic interpolation method DUVMIF of IMSL and obtained identical results. Thus, for a candidate optimal $a^{\prime}$, we pinpoint the optimal $b^{\prime}$ choice using again one dimensional BRENT optimizer and feed $(a^{\prime}, b^{\prime})$ candidate portfolio doubles into bi-dimensional black-box minimization routines described in the text one-by-one. This becomes particularly important for the LINMIN routine as initial bracketing is essential part of minimization routine. As highlighted in \cite{Numerical_Recipes}, ``we would never trade the secure feeling of knowing that a minimum is “in there somewhere” for the dubious reduction of function evaluations that these nonbracketing routines may promise. Please bracket your minima (or, for that matter, your zeros) before isolating them!'' For details, see chapter 10.5.
\item \textbf{Local stage:} If value and the policy functions do not converge within 250 iterations, we switch to local search method within the neighborhood of the obtained candidate $(b^{\prime}, a^{\prime})$ for each grid point of $(b, a, y)$ in the previous step. 
\item \textbf{Stopping rule:} Iterate the procedure defined above for equations (\ref{eq:VR_CE}) to (\ref{eq:q_CE}) until convergence criteria is met. 
\end{enumerate}

With the equilibrium value and price functions as well as decision rules for non-contingent debt borrowing and defaulting, I simulate the model. In particular:
\begin{itemize}
\item Set the number of samples $N =5000$, number of periods $T = 1501$ and $T_0 = 500$. 
\item Use a random number generator to draw sequences of $\varepsilon_t$ and $\psi$ for $t = 1,2,\ldots,T$ to compute the income of the subsequent periods and to evaluate the continuation value of default. It is recommended to keep the draws so that the same values can be used for each sample $n \in N$. 
\item Set the initial endowment $y$ to be mean $y$ and debt holdings $b$ to be zero.
\item Cut the first $T_0$ periods of each sample before computing the moments of the simulation so that randomly chosen initial values will not have any influence on moments.
\end{itemize}

The moments reported in all tables are computed from the 5000 simulated sample paths such that each sample includes 80 periods (20 years) without a default observation. The sample period begins at least 20 periods after having access to the credit markets following a default episode. Default frequency is calculated using the entire simulation. Business cycle moments are reported after HP detrended with a smoothing parameter of 1600.
\end{document}